\documentclass[twocolumn,showpacs,preprintnumbers,amsmath,amssymb,unsortedaddress]{revtex4}

\usepackage{graphicx}  
\usepackage{dcolumn}   
\usepackage{bm}        

\def\lsim{~\rlap{$<$}{\lower 1.0ex\hbox{$\sim$}}}
\def\bsim{~\rlap{$>$}{\lower 1.0ex\hbox{$\sim$}}}

\def\hmpc{\ {\rm {\it h}^{-1}Mpc}}

\def\la{\langle}
\def\ra{\rangle}
\def\dd{{\rm d}}
\def\ln{{\rm ln}}
\def\tr{{\rm tr}}
\def\det{{\rm det}}

\def\mathbi#1{\textbf{\em #1}}

\def\rb{\bar{\rho}_m}

\def\nvh{\hat{\mathbi{n}}}
\def\rvh{\hat{\mathbi{r}}}

\def\rh{\hat{r}}
\def\vk{\mathbi{k}}
\def\vq{\mathbi{q}}
\def\vr{\mathbi{r}}
\def\vx{\mathbi{x}}

\def\vvs{{\bf S}}

\def\vbb{{\rm B}}
\def\vcc{{\rm C}}
\def\vii{{\rm I}}
\def\vmm{{\rm M}}

\def\vrr{{\rm R}}

\def\vxx{{\rm X}}
\def\vyy{{\rm Y}}
\def\grad{\mathbi{$\nabla$}}
\def\epsi{\epsilon}

\def\hw{\hat{W}}
\def\dkk{\Delta_\delta^2(k)}

\def\MNRAS{{Mon.~ Not.~ R.~ Astron.~ Soc.~}}

\def\PRD{{Phys.~ Rev.~ D.~}}
\def\PRL{{Phys.~ Rev.~ Lett.~}}
\def\ApJ{{Astrophys.~ J.~}}
\def\ApJS{{Astrophys.~ J.~ Suppl.~}}

\def\AA{{Astron.~ Astrophys.~}}
\def\AJ{{Astron.~J.}}
\def\Nat{{Nature (London)~}}

\def\AmJP{{Am.~J.~Phys.~}}
\def\JMP{{J.~Math.~Phys.~}}
\def\JCAP{{JCAP}}
\def\JHEP{{Journal~of~High~Energy~Physics}}

\begin{document}

\title{Statistical properties of the linear tidal shear}

\author{Vincent Desjacques and Robert E. Smith
\small \\ \vspace{0.2cm} Institute for Theoretical Physics, 
University of Z\"urich, Winterthurerstrasse 190, CH-8057 Z\"urich, 
Switzerland \\
email: {\tt dvince@physik.uzh.ch}, {\tt res@physik.uzh.ch}}


\begin{abstract}

Large-scale structures originate from coherent motions induced by
inhomogeneities in the primeval gravitational potential. Here, we
investigate the two-point statistics of the second derivative of the
potential, the tidal shear, under the assumption of Gaussianity.  We
derive an exact closed form expression for the angular averaged,
two-point distribution of the shear components which is valid for an
arbitrary Lagrangian separation. This result is used to write down
the two-point statistics of the shear eigenvalues in compact form.
Next, we examine the large-scale asymptotics of the correlation of
the shear eigenvalues and the alignment of the principal axes.  The
analytic results are in good agreement with measurements  obtained
from random realizations of the gravitational potential.  Finally, we
show that a number of two-point distributions of the  shear
eigenvalues are well approximated by Gaussian bivariates  over a wide
range of separation and smoothing scales. We speculate  that the
Gaussian approximation also holds for multiple point  distributions of
the shear eigenvalues. It is hoped that these results will be relevant
for studies aimed at describing the properties of the  (evolved)
matter distribution in terms of the statistics of the  primordial
shear field.

\end{abstract}

\maketitle

\section{Introduction}
\label{sec:intro}

In the currently favored $\Lambda$CDM cosmology, the galaxies and the
large-scale structures we observe today are thought  to arise from the
hierarchical growth of initially tiny  Gaussian fluctuations. Galaxy
surveys~\cite{galaxysurvey} reveal that large-scale  structures on
scale $\sim 10-100\hmpc$ form a predominantly filamentary network
whose principal constituents - clusters, filaments and walls -
accumulate on the  boundary of large voids. Numerical simulations and
semianalytic approaches  have been very  successful in reproducing the
observed matter distribution~\cite{simulations}, whereas   analytic
models  based on the spherical infall ~\cite{GunnGott1972} predict
mass  functions, merging histories and spatial clustering of bound
objects that are in reasonable agreement with the observations
~\cite{analytics}.  In parallel, several methods have been proposed to
describe quantitatively the structures observed both in the Universe
and in the simulations.  The full hierarchy of correlation functions,
the two-point statistics in particular~\cite{Peebles1980}, remains the
most widely used statistical tools to distinguish between different
scenarios of structure formation and constrain the cosmological
parameters.  Topological estimators such as Minkowski functionals
~\cite{topologics} provide useful complementary information on the
morphological characteristics of the filamentary network. Furthermore,
various identification  algorithms abstracting the spatial patterns in
points, lines, etc. have been proposed in an attempt to improve upon
current topological measures~\cite{algorithms}.

Although the sequence in which large-scale structures form is still a
matter of debate, many lines of evidence suggest that the filamentary
pattern seen in observations and in N-body simulations is a
consequence of the spatial coherence of the initial tidal shear
~\cite{Bondetal1996}. While the spherical infall
model~\cite{GunnGott1972} captures the essential features of
gravitationally induced collapse,  the primeval shear field  has also
been shown to play a crucial role in the   formation of nonlinear
structures ~\cite{shearfield}. As demonstrated
in~\cite{ShethTormen2002}, the inclusion of nonsphericity in the
collapse dynamics yields a  better fit to the halo mass functions
measured in N-body simulations. Yet another important manifestation of
the tidal shear is the alignment of shape and angular momentum of
objects~\cite{Binggeli1982,Argyresetal1986,West1989,Westetal1989,
Catelanetal2001}. Numerical studies of the $\Lambda$CDM cosmology
report strong correlations in the alignment of galaxies, haloes,
massive clusters, or voids~\cite{Heavensetal2000,CroftMetzler2000,
Hopkinsetal2005,Hahnetal2007,Platenetal2008}, reflecting the coherence
of the matter distribution out to large distances.

In the ``Cosmic Web'' picture outlined in~\cite{Bondetal1996}, the
correspondence between structures in the evolved density field and
local properties of the linear tidal  shear should, in principle,
allow us to estimate the morphology of the matter distribution. In
practice however, this correspondence has not been much exploited
principally because of the lack of theoretical results. In spite of
the progress made in the analysis of Gaussian random fields
~\cite{Adler1981,PeacockHeavens1985,Bardeenetal1986} applied to the
formation of large-scale structures, the statistics of  the shear has
received little attention. Doroshkevich~\cite{Doroshkevich1970} first
calculated the probability  distribution of the shear eigenvalues and
ascertained the amount of material being incorporated in sheetlike
structures or pancakes. Reference \cite{DoroshkevichShandarin1978}
reexamined the formation of these pancakes and derived a distribution
function for the largest eigenvalue of the shear tensor. Reference
\cite{LeeShandarin1998} computed conditional probability distributions
for individual shear eigenvalues and obtained an analytic
approximation to the halo mass function. Also,
\cite{CatelanPorciani2001} explored the two-point correlation of the
tidal shear components, but they did not discuss probability
distributions.

In this paper, we carry out the analysis of the 2-point statistics
of the linear tidal shear at two distinct (Lagrangian) positions.  We
extend the study of ~\cite{Desjacques2007}, who derived an
expression for the shear 2-point statistics smoothed at two different
scales, but evaluated at a single position.  This work is essentially
intended to provide theoretical results  that could improve the 
statistical description of the tidal shear and, therefore, of the 
Cosmic Web.
Section~\ref{sec:2pdf} is devoted to the calculation of
the 2-point distributions. A careful examination of  the small and
large scale behavior of the 2-point distribution of  shear components
suggests a compact expression for the joint  distribution of shear
eigenvalues. This result is used in Sec.~\ref{sec:correl} to explore the
asymptotic behavior  of the correlations of shear eigenvalues and
principal axes. In Sec.~\ref{sec:gaussian}, conditional 2-point
distributions obtained from random realization of the linear shear
field are compared  with theoretical Gaussian distributions. It is
argued that Gaussian  multivariates provide a good description of the
$n$-point distributions of shear eigenvalues at all separation and
smoothing length. This suggests the possibility of implementing
nonspherical collapse in current analytic models of structure
formation using well-known results of the theory of Gaussian random 
fields.

\section{Shear}
\label{sec:basics}

The comoving Eulerian position of a particle can be generally expressed  
as a mapping
\begin{equation}
\vx=\vq+\vvs(\vq,a)\;,
\label{eq:za} 
\end{equation}
where $\vq$ is the Lagrangian (initial) position, $a$ is the scale 
factor and $\vvs$ is the
displacement field. Continuity implies  that the Eulerian density
contrast $\delta(\vx,a)$ is given by the reciprocal of the Jacobian
for the transformation (\ref{eq:za}), $1+\delta(\vx,a)=|{\rm
det}\left(\partial\vq/\partial\vx\right)|$.  Singularities occur in
this mapping whenever at least one of the eigenvalues  is positive,
signaling crossing of particle trajectories at that Eulerian
point. The initial deformation tensor or strain field ${\rm
D}_{ij}=\partial_i\vvs_j$ ($\partial_i\equiv \partial/\partial q_i$)
thus encodes important  information on the collapse of large-scale
structures.

In the Zeldovich approximation, the displacement
field is $\vvs(q,a)=-D_+(a)\grad\Phi(\vq)$, where
$\Phi(\vq)=\phi(\vq,a)/4\pi G\rb(a)a^2 D_+(a)$ is the perturbation
potential, $\phi(\vq,a)$ is the Newtonian gravitational potential,
$\rb$ is the average matter density and $D_+(a)$ is the linear growth
factor (normalized such that $D_+(a)=1$)~\cite{Peebles1980}. The 
strain tensor now is proportional to the second derivatives of the 
perturbation potential. For convenience, we will introduce the real, 
symmetric tensor
\begin{equation}
\xi_{ij}(\vq)=\frac{1}{\sigma}\, \frac{\partial^2\Phi}{\partial
q_i\partial q_j}(\vq)\;,
\label{eq:strain}
\end{equation}
where $\sigma$ is the root-mean-square (rms) variance of density 
fluctuations
$\delta(\vq)\equiv\Delta\Phi(\vq)$ linearly extrapolated to the
present time. We shall henceforth refer to $\xi_{ij}$ as the shear
tensor. Notice that $\xi_{ij}$ is dimensionless while $\Phi(\vq)$ has
the unit of (length)$^2$. We will also assume that these fields are
smoothed at some characteristic scale $R_S$ with a spherically
symmetric window $W(r,R_S)$. Although many choices   are possible
for such a filtering function, we will adopt a top-hat filter
throughout this paper, so that the variances are related to spherical
volumes of radius $R_S$.

Let $\Lambda={\rm diag(\lambda_1,\lambda_2,\lambda_3})$ be the
diagonalized shear tensor. For Gaussian initial conditions, the
1-point probability distribution of  the shear eigenvalues
derived in~\cite{Doroshkevich1970} can be written as
\begin{equation}
P(\lambda_1,\lambda_2,\lambda_3)=
\frac{15^3}{8\pi\sqrt{5}}\,e^{-Q_1(\Lambda,\Lambda)}\Delta(\lambda)\;,
\end{equation}
where, for shorthand convenience, 
\begin{equation}
Q_1(\vxx,\vyy)=\frac{3}{4}\left[5\tr\left(\vxx\vyy\right)-
\left(\tr\vxx\right)\left(\tr\vyy\right)\right]
\end{equation}
is some (indefinite) quadratic form over the space of real matrices,
and 
\begin{equation}
\Delta(\lambda)={\rm det}\left(\lambda_i^{3-j}\right)= \prod_{i<j}
\left(\lambda_i-\lambda_j\right)
\label{eq:vandermonde}
\end{equation}
is the Vandermonde determinant in the arguments
$\lambda_1,\lambda_2,\lambda_3$. Our definition (\ref{eq:strain}) of
the  shear tensor makes the probability
$P(\lambda_1,\lambda_2,\lambda_3)$  independent of the filtering scale
$R_S$. For instance, one finds that, for ambient field points, the
probability   of all three eigenvalues being positive is
$P(+++)=0.08$, and that of  the configurations $(++-)$ and $(+--)$ is
0.84.  Note, however, that these values depend strongly on the peak 
height, $\nu=\delta/\sigma$, of the region under
consideration, the highest density peaks being predominantly
spherical~\cite{Bernardeau1994,Pogosyanetal1998}.

\section{Two-point statistics}
\label{sec:2pdf}

Desjacques~\cite{Desjacques2007} extended the results of
~\cite{Doroshkevich1970} to the joint statistics  of the shear
smoothed at different scales. However, he confined his  calculation to
the case in which the joint distributions are evaluated  at a single
Lagrangian position.  Here we address the general case and calculate
the joint distribution  of the shear components $\xi_{ij}(\vq_1)$ and
$\xi_{kl}(\vq_2)$ for arbitrary Lagrangian separations
$\vr=\vq_2-\vq_1$.  We shall assume throughout this paper that the
initial fluctuations are  described by the Gaussian statistics. This
assumption is  remarkably well supported by the latest measurements of
the cosmic microwave background (CMB)~\cite{gaussianity}.

\subsection{Shear correlations}

We take the components $\xi_{ij}(\vq_1)$ and $\xi_{kl}(\vq_2)$ to be 
smoothed at two different (comoving) scales $R_1$ and $R_2$, 
respectively. The spectral parameter
\begin{equation}
\gamma\equiv \frac{1}{\sigma_1\sigma_2} \int_0^\infty\!\!\dd\ln
k\,\dkk\, \hw(k,R_1)\hw(k,R_2)\;,
\label{eq:xi01}
\end{equation}
$0\leq\gamma\leq 1$, is a measure of the correlation between these
scales.  Here, $\dkk\equiv k^3 P_\delta(k)/2\pi^2$ is the
dimensionless  power spectrum of the density field, $\hw(k,R_i)$ is
the Fourier transform of $W(r,R_i)$, and $\sigma_i$ is the  rms
variance of  density fluctuations $\delta(\vq)$ smoothed at  scale
$R_i$. 

Statistical isotropy and symmetry imply  that, in position space, 
the 2-point correlation functions of an arbitrary symmetric tensor 
field $\xi_{ij}(\vq)$ must be of the form
\begin{eqnarray}
\lefteqn{\la \xi_{ij}(\vq_1)\xi_{lm}(\vq_2)\ra = \Psi_1(r)\,\rh_i
\rh_j \rh_l \rh_m} \nonumber \\ && +\,\Psi_2(r)\left(\rh_i
\rh_l\delta_{jm}+\rh_i \rh_m\delta_{jl} +\rh_j \rh_l\delta_{im}+\rh_j
\rh_m\delta_{il}\right) \nonumber \\ && +\,\Psi_3(r)\left(\rh_i
\rh_j\delta_{lm}+\rh_l \rh_m\delta_{ij}
\right)+\Psi_4(r)\,\delta_{ij}\delta_{lm} \nonumber \\ &&
+\,\Psi_5(r)\left(\delta_{il}\delta_{jm}+\delta_{im}\delta_{jl}
\right) \;,
\label{eq:correl}
\end{eqnarray}
where $r=|\vq_1-\vq_1|$, $\rh_i=r_i/r$ is a unit vector and the
functions $\Psi_i(r)$ depend  on $r$ only.  This is the most general
ansatz for the isotropic sector of the fourth order correlation
function $\la \xi_{ij}\xi_{lm}\ra(\vr)$. Symmetry requires that $\rh$
appears in even number pairs.  In the case of a scalar (spin-0) tensor
such as the linear tidal shear defined in Eq.~(\ref{eq:strain}),
$\Psi_2=\Psi_3$ and  $\Psi_4=\Psi_5$. Notice that
Eq.~(\ref{eq:correl})  holds regardless of the statistical
properties of the gravitational potential.  However, when $\Phi(\vq)$
is Gaussian, the functions $\Psi_i$ may be conveniently expressed as
\begin{eqnarray}
\Psi_1(r)\!\!\! &=& \!\!\! \int_0^\infty\!\!\dd\ln
k\,\Delta^2(k) j_4(kr) \\ \Psi_3(r) \!\!\! &=& \!\!\! 
\int_0^\infty\!\!\dd\ln k\,\Delta^2(k)\left[-\frac{1}{7}j_2(kr)-
\frac{1}{7}j_4(kr)\right] \nonumber \\ \Psi_5(r) \!\!\! &=& \!\!\! 
\int_0^\infty\!\!\dd\ln k\,\Delta^2(k)\left[\frac{1}{15}j_0(kr)+
\frac{2}{21}j_2(kr)+\frac{1}{35}j_4(kr)\right] \nonumber
\label{eq:psif}
\end{eqnarray}
where $\Delta^2(k)\equiv \dkk\hw_1\hw_2/(\sigma_1\sigma_2)$ and
$j_\ell(x)$ are spherical Bessel functions of the first kind. The
$\Psi_i$ can be equivalently expressed in terms of the auxiliary 
functions $J_n\equiv n r^{-n}\int_0^r\!\!\dd s\,\psi(s)s^{n-1}$
~\cite{LeePen2001,Crittendenetal2001,CatelanPorciani2001}, where
\begin{equation}
\psi(r)=\Psi_1(r)+10\Psi_3(r)+15\Psi_5(r)
\label{eq:corrdens}
\end{equation}
is the cross correlation between the density enhancement 
$\delta/\sigma$
smoothed at two different scales. In the limit $r\rightarrow 0$, 
both $\Psi_1$ and $\Psi_3$ vanish while  $\Psi_5$ tends towards
$\gamma/15$, so that $\psi\rightarrow\gamma$.

\subsection{Two-point probability distribution}

Owing to the symmetry of $\xi_{ij}$, only six components of the shear
are independent. Following the notation of~\cite{Bardeenetal1986}, let
$\tilde{\xi}=\{\tilde{\xi}_A,A=1,\dots,6\}$ designate the
six-dimensional  vector whose components are equal to the components
$ij=11,22,33,12,13,23$ of the shear tensor.
The joint probability
distribution $P\left(\xi_1,\xi_2;\vr\right)$ of the  shear tensor
$\xi_1=\xi_{ij}(\vq_1)$ and $\xi_2=\xi_{ij}(\vq_2)$ is given by a 
multivariate Gaussian whose
covariance matrix $\vcc$ has 12 dimensions.  This $12 \times 12$
matrix may be partitioned into four $6\times 6$  block matrices,
$\vcc_1=\la\tilde{\xi}_1\tilde{\xi}_1^\top\ra$ in  the top left
corner, $\vcc_2=\la\tilde{\xi}_2\tilde{\xi}_2^\top\ra$ in the bottom
right corner, $\vbb=\la\tilde{\xi}_1\tilde{\xi}_2^\top\ra$ and its
transpose in the  bottom left and top right corners, respectively,
where
\begin{equation}
\vcc_1=\vcc_2=\left(\begin{array}{cc}\vmm_1/15 & 0 \\  0 &
\vii/15\end{array}\right),~~~ \vmm_1=\left(\begin{array}{ccc} 3 & 1  &
1 \\ 1 & 3 & 1 \\ 1 & 1 & 3\end{array}\right) \;,
\end{equation} 
and $\vii$ is the $3\times 3$ identity matrix. 

Unlike $\vcc_1$ and
$\vcc_2$, the cross correlation matrix $\vbb$ generally is a function 
of the separation $\vr$. Using the harmonic decomposition of the tensor 
products $\rvh\otimes\dots\otimes\rvh$ which appear in 
Eq.~(\ref{eq:correl}), $\vbb(\vr)$ can be written as follows~:
\begin{equation}
\vbb(\vr)=\left(\begin{array}{cc}\vbb_1(\vr) & \vbb_3(\vr) \\  
\vbb_3(\vr) & \vbb_2(\vr) \end{array}\right)\;,
\end{equation}
with $3\times 3$ block matrices
\begin{eqnarray}
\vbb_1(\vr)&=&\frac{1}{15}\psi(r)\,\vmm_1+\sum_{\ell=2,4}
\vbb_1^{\ell,m}(r)\,Y_\ell^m(\rvh) \nonumber \\
\vbb_2(\vr)&=&\frac{1}{15}\psi(r)\,\vii+\sum_{\ell=2,4}
\vbb_2^{\ell,m}(r)\,Y_\ell^m(\rvh) \nonumber \\
\vbb_3(\vr)&=&\sum_{\ell=2,4}\vbb_3^{\ell,m}(r)\,Y_\ell^m(\rvh)\;.
\end{eqnarray}
$Y_\ell^m(\rvh)$ are spherical harmonics and $\vbb_i^{\ell,m}(r)$ are
$3\times 3$ matrices which satisfy $(\vbb_i^{\ell,m})^\dagger=(-1)^m
\vbb_i^{\ell,m}$ ($\vbb_i(\vr)$ are real-valued matrix). An explicit
calculation of these matrices is not necessary as we will focus on the
contribution of the monopole terms.  Again, symmetry implies that
only the harmonics with multipoles $\ell=0$, 2, 4 and even $m$ appear
in the decomposition.  Furthermore, it is worth noticing that the
joint  probability density  $P(\xi_1,\xi_2;\vr)$ is invariant under
any arbitrary rotation of the  coordinate system~\cite{isotropic},
\begin{equation}
P(\xi_1',\xi_2';\vr')=P(\vrr\xi_1\vrr^\top,\vrr\xi_2\vrr^\top,
\vrr^\top\vr)=P(\xi_1,\xi_2,\vr)\;,
\label{eq:changecoord}
\end{equation}
where primes denote quantities in the new coordinate frame. However,
in a given coordinate system, its functional form changes when $\rvh$
moves over the unit sphere. Namely, the transverse and longitudinal 
components of the 2-point shear correlation vary with the 
orientation of the separation vector. Therefore, the anisotropic 
structure of $\la \xi_{ij}\xi_{lm}\ra(\vr)$ should come as no surprise
~\cite{CatelanPorciani2001}.

\subsection{Angle average distribution}
\label{sec:angleaverage}

We are primarily interested in the angular average of the 2-point 
probability distribution of the linear tidal shear, 
\begin{equation}
P(\xi_1,\xi_2;r)=\frac{1}{4\pi}\int\!\!\dd\Omega_{\rvh}\,
P(\xi_1,\xi_2;\vr)\;,
\label{eq:avangle}
\end{equation}
which is a function of the separation $r$ solely. To get insight on
the functional form of $P(\xi_1,\xi_2;r)$,  we will examine the
small-scale ($r\ll 1$) and large-distance asymptotic ($r\gg 1$)
behavior of the probability density $P(\xi_1,\xi_2,\vr)$.  In both
regimes, the cross correlation matrix $\vbb$ can be expressed as
$\vbb(\vr)=\bar{\vbb}+\delta\vbb(\vr)$, where $\delta\vbb$  is a small
perturbation and $\bar{\vbb}$ is the zero order contribution, which 
is either $\bar{\vbb}=\gamma\vcc_1$ (when $r\rightarrow 0$) or
$\bar{\vbb}=0$ (when $r\rightarrow \infty$). The quadratic form  which
appears in the probability distribution $P(\xi_1,\xi_2;\vr)$,
\begin{equation}
P(\xi_1,\xi_2;\vr)=\frac{1}{\left(2\pi\right)^6|\det\vcc|^{1/2}}\,
e^{-Q(\xi_1,\xi_2;\vr)}\;,
\end{equation}
can be computed easily using Schur's identities. Here, $\det\vcc$ is
the determinant of the covariance matrix $\vcc$. Expanding the
exponential  in the perturbation $\delta\vbb$ and averaging over
directions  $\rvh$, we find
\begin{eqnarray}
\lefteqn{\frac{1}{4\pi}\int\!\!\dd\Omega_{\rvh}\exp\left[-Q(\xi_1,\xi_2;
\vr)\right]} && \nonumber \\ 
&& \approx \left\{1-2\frac{\left(\psi-\gamma\right)}{1-\gamma^2}\,
\left[\frac{\gamma}{1-\gamma^2}\,\left(Q_1(\xi_1,\xi_1)+Q_1(\xi_2,\xi_2)
\right)\right.\right. \nonumber \\ && \left.\left.\frac{}{}\!\! 
-Q_1(\xi_1,\xi_2)
\right]\right\}\,e^{-\left[Q_1(\xi_1,\xi_1)+Q_1(\xi_2,\xi_2)-2\gamma 
Q_1(\xi_1,\xi_2)\right]/\left(1-\gamma^2\right)} \nonumber \\ 
&& \hspace{6cm}\mbox{($r\ll 1$)} \;, \\
\label{eq:probsmallscale}
&&\approx \left[1+2\,\psi\,Q_1(\xi_1,\xi_2)\right]\,
e^{-Q_1(\xi_1,\xi_1)-Q_1(\xi_2,\xi_2)} \nonumber \\
&& \hspace{6cm}\mbox{($r\gg 1$)}\;,
\label{eq:problargescale}
\end{eqnarray}
to first order in $\delta\vbb$. Interestingly, these perturbative 
expansions precisely match the small-scale and asymptotic large $r$ 
behavior of the function $\exp[-Q_2(\xi_1,\xi_2,r)]$, where
\begin{equation}
Q_2(\xi_1,\xi_2;r)=\frac{Q_1(\xi_1,\xi_1)+Q_1(\xi_2,\xi_2)
-2\,\psi\,Q_1(\xi_1,\xi_2)}{1-\psi^2}\;.
\end{equation}
This strongly suggests that the 2-point probability distribution 
$P(\xi_1,\xi_2;r)$ of the shear tensor may be written exactly as
\begin{equation}
P(\xi_1,\xi_2;r)=\frac{1}{20}\left(\frac{15}{2\pi}\right)^6
\left(1-\psi^2\right)^{-3}\,e^{-Q_2(\xi_1,\xi_2;r)}\;.
\label{eq:2pdfc}
\end{equation}
This distribution depends on the separation $r$ through the density
correlation $\psi(r)$ only. Gaussianity and invariance  under rotation
requires that $P$ be a function of $\tr(\xi_i^2)$, $(\tr\xi_i)^2$,
$\tr(\xi_1\xi_2)$ and $\tr\xi_1\tr\xi_2$ solely.  Although we have not
been able to rigorously prove that Eq.~(\ref{eq:2pdfc}) correctly
describes the 2-point distribution in the intermediate region, we
have found that it agrees with the result of a direct numerical
integration of Eq.~(\ref{eq:avangle}) (for various choices of $\xi_1$,
$\xi_2$ and $r$) up to the numerical
accuracy. Furthermore, the fact that, in the limit $r\ll 1$,
Eq.~(\ref{eq:2pdfc}) reduces to the joint probability density derived
in~\cite{Desjacques2007} is another indication of correctness.

\subsection{Joint distribution of the eigenvalues}

We now choose a  coordinate system such that the coordinate axes are
aligned with the  principal axes of $\xi_1$. Let $\Lambda_1$ and
$\Lambda_2$ be the diagonal  matrices consisting of the three ordered
eigenvalues $x_1\geq x_2\geq x_3$ and $y_1\geq y_2\geq y_3$ 
of the deformation tensors
$\xi_1$ and $\xi_2$, respectively. The principal axes are now labeled
according to this ordering. With this choice of coordinate,
$\xi_1=\Lambda_1$ and $\xi_2=\vrr\Lambda_2\vrr^\top$, where $\vrr$ is
an orthogonal matrix that defines the orientation of the eigenvectors
of $\xi_2$ relative to those of $\xi_1$. To preserve the orientation
of the principal axis frames, we further impose the condition that the
determinant of $\vrr$ must be +1. Namely, $\vrr$ belongs to the
special  orthogonal group SO(3).  The properties of the trace imply
that $Q_1(\xi_2,\xi_2)=Q_1(\Lambda_2,\Lambda_2)$, while the term
$Q_1(\xi_1,\xi_2)=Q_1(\Lambda_1,\vrr\Lambda_2\vrr^\top)$ depends  on
the rotation matrix.

\begin{figure}
\center \resizebox{0.45\textwidth}{!}{\includegraphics{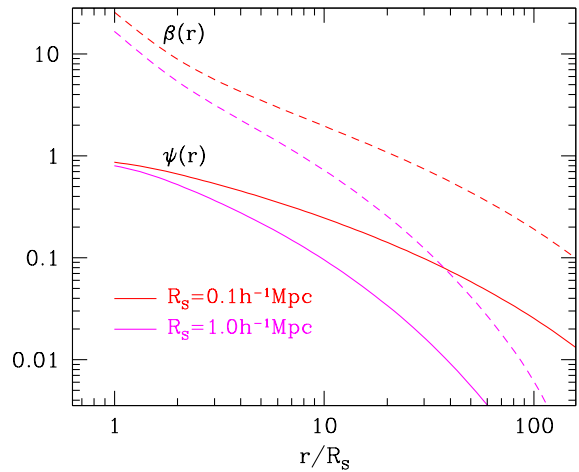}}
\caption{The density correlation $\psi(r)$ and the parameter $\beta(r)$
(cf. text) as a function of the ratio $r/R_S$ for two different smoothing 
lengths $R_S=0.1$ and 1$\hmpc$.}
\label{fig:gam}
\end{figure}

To obtain the joint probability distribution of  the ordered eigenvalues 
of the shear tensor, ``angular'' variables, such as those appearing in  
$Q_1(\xi_1,\xi_2)$, have to be integrated over.
The volume measure $\dd\xi$ for the space of real $3\times 3$
symmetric  matrices can be expressed in terms of the nonincreasing
sequence of  eigenvalues $\lambda_i$ ($=x_i$ or $y_i$) as
\begin{equation}
\dd\xi=8\pi^2\,\Delta(\lambda)\dd^3\lambda\,\dd\vrr\;.
\label{eq:vol1}
\end{equation}
Here, $\dd\vrr$ is the Haar measure on the group SO(3) normalized to
$\int\dd\vrr=1$, $\dd^3\lambda=\dd\lambda_1\dd\lambda_2\dd\lambda_3$
and  $\Delta(\lambda)$ is the Vandermonde determinant
equation~(\ref{eq:vandermonde}).  When the rotation matrices $\vrr$ are
parametrized in terms of the  Euler angles $0\leq \varphi,\psi\leq
2\pi$, $0\leq\vartheta\leq\pi$, the Haar measure takes the familiar
form
\begin{equation}
\dd\vrr=\frac{1}{8\pi^2}\sin\vartheta\,\dd\varphi\dd\vartheta\dd\psi\;.
\label{eq:haar}
\end{equation}
Since the quadratic form $Q$ depends only on the relative orientation 
of the eigenvector triads of $\xi_1$ and $\xi_2$, we can immediately
integrate over  one of the SO(3) manifolds.  The relevant volume is
$8\pi^2/4=2\pi^2$. The factor 4 comes from not  caring whether the
rotated axis points in the positive or negative
direction~\cite{Bardeenetal1986}.  The essential problem is the
calculation of the integral over the manifold that defines relative,
distinct triad orientations,
\begin{equation}
{\cal I}=\int_{\rm SO(3)}\!\!\!\!\!\!\dd\vrr\,\exp\left[\beta\,
\tr\left(\vrr\Lambda_2\vrr^\top\!\Lambda_1\right)\right]\;,
\label{eq:intso3}
\end{equation}
where we have defined $\beta(\psi)=(15/2)\psi/(1-\psi^2)$.  

There is no analytic, closed-form solution to this integral, although
an  exact determinantal expression was derived when averaging over the
unitary group U(N)~\cite{Harish1958,ItzyksonZuber1980}.  An asymptotic
representation can be obtained when $\beta\gg 1$. For  reasonable
values of $R_S$, this occurs when the  separation $r$ is less than a
few smoothing radii (see Fig.~\ref{fig:gam}). In general, the
integral~(\ref{eq:intso3}) can be expressed as a hypergeometric
series with the symmetric 3$\times$3 matrices $\beta\Lambda_1$ and
$\Lambda_2$ as argument (see Appendix~\S\ref{app:orientation}). In 
the notation of ~\cite{Muirhead1982}, ${\cal I}\equiv {_0F_0^{(3)}}$ 
where, at second order in $\beta$,
\begin{eqnarray}
\lefteqn{_0F_0^{(3)}\!\left(\beta\Lambda_1,\Lambda_2 \right)=
1+\frac{\beta}{3}\tr\Lambda_1\,\tr\Lambda_2+\frac{\beta^2}{18}
(\tr\Lambda_1)^2(\tr\Lambda_2)^2} && \nonumber \\
&& +\frac{\beta^2}{90}\left[3\tr(\Lambda_1^2)-(\tr\Lambda_1)^2\right]\,
\left[3\tr(\Lambda_2^2)-(\tr\Lambda_2)^2\right] \;.
\label{eq:intso3form1}
\end{eqnarray}
Higher order terms are intentionally omitted as they are not used in 
the present analysis.

A straightforward calculation shows that the eigenvalue joint probability 
distribution $P(x,y;r)$ evaluates to
\begin{eqnarray}
P(x,y;r) &=& \frac{15^6}{320\pi^2}\left(1-\psi^2\right)^{-3}\, 
_0F_0^{(3)}\!(\beta\Lambda_1,\Lambda_2) \nonumber \\  && \times
e^{-Q_{12}(\Lambda_1,\Lambda_2;r)}\,\Delta(x)\Delta(y)\;,
\label{eq:2pdfe}
\end{eqnarray}
where 
\begin{equation}
Q_{12}=\frac{Q_1(\Lambda_1,\Lambda_1)+Q_1(\Lambda_2,\Lambda_2)
+\frac{3}{2}\psi\left(\tr\Lambda_1\right)\left(\tr\Lambda_2\right)}
{\left(1-\psi^2\right)}\;.
\label{eq:Q12}
\end{equation}
Notice that, in the limit $|\psi|\ll 1$, the joint probability
distribution  $P(x,y;r)$ tends, as it should be, toward the product
of the  individual 1-point probability distribution.   Using Bayes'
theorem, we easily derive a conditional distribution $P(x|y;r)$ for
the shear eigenvalues $x_i$ given the $y_i$s and a separation $r$,
\begin{equation}
P(x|y;r)=\frac{15^3}{8\pi\sqrt{5}}\left(1-\psi^2\right)^{-3}\, 
_0F_0^{(3)}\!(\beta\Lambda_1,\Lambda_2)e^{-Q_{1|2}}\,\Delta(x)\;,
\label{eq:conpdf}
\end{equation}
where the quadratic form $Q_{1|2}(\Lambda_1,\Lambda_2;r)$ is
\begin{equation}
Q_{1|2}=\frac{Q_1(\Lambda_1,\Lambda_1)+\psi^2\,Q_1(\Lambda_2,\Lambda_2)
+\frac{3}{2}\psi(\tr\Lambda_1)(\tr\Lambda_2)}{1-\psi^2}\;.
\label{eq:Q1|2}
\end{equation}
A direct numerical integration convinced us that the probability
distribution (\ref{eq:conpdf}) is normalized to unity (we used the
multidimensional integrator {\small DCUHRE} described in
~\cite{Berntsenetal1991}). We believe that Eqs.~(\ref{eq:2pdfe}) and
(\ref{eq:conpdf}) are exact expressions for the 2-point and
conditional  probability distribution function of the shear
eigenvalues. They generalize the results obtained in
~\cite{Doroshkevich1970,Desjacques2007}.

\section{Asymptotics of correlation functions}
\label{sec:correl}

Instead of attempting a brute force calculation of the correlation
functions through a direct integration of the probability density
~(\ref{eq:2pdfe}), we will examine the large-scale asymptotic behavior
solely. We will nonetheless infer analytic approximations to the
correlations  of shear eigenvalues which are accurate on all scales.

\subsection{Eigenvalues}

In order to derive the correlation function for the eigenvalues of the
shear tensor in the asymptotic regime $r\gg 1$  for which $\psi\ll 1$
and $\beta(\psi)\ll 1$, we transform to the new set of variables
$\left\{\nu_i,e_i,p_i,i=1,2\right\}$, where $\nu_1=x_1+x_2+x_3$,
$e_1=(x_1-x_3)/2\nu_1$ and  $p_1=(x_1-2 x_2+x_3)/2\nu_1$.  The
variables $(\nu_2,e_2,p_2)$ are defined as a function of the
eigenvalues $y_i$ in a similar way.  $e_i$ and $p_i$ are the shear
ellipticity and prolateness, respectively.  Our choice of  ordering
impose the constraints $e_i\geq 0$ and  $-e_i\leq p_i\leq e_i$.

The cross correlation function $\zeta_{ij}(r)$, or 2-point connected 
moment of the shear eigenvalues $x_i$ and $y_j$ is defined as
\begin{equation}
\la x_i\ra\la y_j\ra+\zeta_{ij}(r)=
\int\!\!\dd^3 x\dd^3 y\,P\left(x,y;r\right)\,x_i y_j\;. 
\label{eq:ceig1}
\end{equation}
The integration over the variables $p_i$ and $e_i$ is straightforward
to second order in $\beta$. In this respect, notice that the volume 
measure $\dd^3x$ and the Vandermonde determinant $\Delta(x)$ are  
$\dd^3x=(2/3)\nu_1^2\,\dd\nu_1\dd e_1\dd p_1$  and
$\Delta(x) = 2\nu_1^3\, e_1\left(e_1^2-p_1^2\right)$,
respectively.  Furthermore, with the help of the series expansion
~(\ref{eq:intso3form1}) about $\beta=0$, the average over the
relative orientation can be reduced to
\begin{equation}
_0F_0^{(3)}\approx 1+\frac{\beta}{3}\nu_1\nu_2+\frac{\beta^2}{9}\nu_1^2
\nu_2^2
\left[\frac{1}{2}+\frac{2}{5}\left(3 e_1^2+p_1^2\right)
\left(3 e_2^2+p_2^2\right)\right]
\end{equation}
in the aforementioned coordinates. The integration over the peak heights 
$\nu_1$ and $\nu_2$ is then easily performed and gives
\begin{equation}
\zeta_{ii}(r) = \frac{1}{9}\,\psi(r)+\frac{1}{10}\la x_i\ra^2\,\psi^2(r)
\label{eq:ceig2}
\end{equation} 
at second order for all three eigenvalues. To derive this result, we 
have used the following expressions for the mean of the shear  
eigenvalues~\cite{Doroshkevich1970,LeeShandarin1998},
\begin{equation}
\la x_1\ra=\la x_3\ra=\frac{3}{\sqrt{10\pi}},~~\la x_2\ra=0\;.
\label{eq:meaneig}
\end{equation}
Hence, there is no second order contribution to $\zeta_{22}(r)$. It is 
worth remembering that the variance $\la x_i^2\ra$ of each 
eigenvalue is
\begin{equation}
\la x_1^2\ra=\la x_3^2\ra=\frac{13\pi-27}{30\pi},~~
\la x_2^2\ra=\frac{2}{15}\;,
\label{eq:vareig}
\end{equation}
close to the value of $1/9$. This suggests that the correlation
functions $\zeta_{ii}(r)$ are well approximated by $\psi(r)/9$ on all
scales. This is not entirely surprising since, in the case where the 
correlations between eigenvalues are equal, the constraint
$\sum_{i,j}\zeta_{ij}(r)=\psi(r)$ would imply $\zeta_{ij}(r)=\psi(r)/9$.

\begin{figure}
\center \resizebox{0.45\textwidth}{!}{\includegraphics{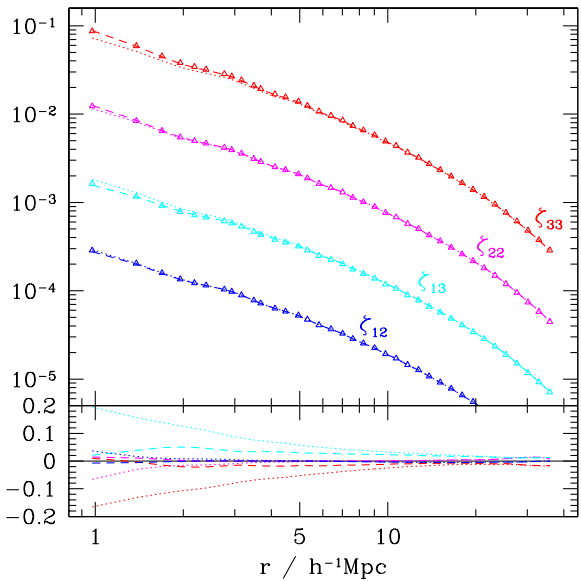}}
\caption{A comparison between a measurement of the auto- and
cross-correlation of the shear eigenvalues from realizations of
Gaussian  random fields (symbols) and our analytic approximations
(curves). All the correlations but $\zeta_{33}$ have been shifted
vertically for clarity.  On large scale, the correlations  asymptote
to the exact result $\psi(r)/9$ shown as the dotted curve.  On small
scale, however, $\bar{\zeta}_{ij}(r)$ [as defined in
Eqs.~\ref{eq:ceig2} and \ref{eq:ceig4}]  provides a better fit to the
measurement (dashed curve).  The bottom panel shows the
fractional error. Notice that the density correlation $\psi(r)$ used
for the comparison is calculated from the numerical realizations to
account for the missing power at small and large scales.}
\label{fig:corr1}
\end{figure}

In  order to verify this assumption, we have generated random
realizations of the  potential field $\Phi(\vq)$ for the $\Lambda$CDM
cosmology considered here on a  256$^3$ mesh of size 250$\hmpc$. The
eigenvalues of the shear tensor are computed on the same grid using
standard FFT (Fast Fourier Transform) techniques. More precisely, the
Fourier modes of the shear  are computed using the relation
$\xi_{ij}(\vk)=k_i k_j\Phi(\vk)/\sigma$.  Once $\xi_{ij}(\vk)$ is
Fourier-transformed back, the shear eigenvalues  $x_i$ as well as the
density $\nu=x_1+x_2+x_3$ are calculated at  each grid point. Lastly,
after having checked that the measured variances $\la x_i^2\ra$ match
well the analytic expectation~(\ref{eq:vareig}), we calculate
correlation functions of the shear eigenvalues. Note  that the
gravitational potential is smoothed on scale $R_S=1\hmpc$.

In Fig.~\ref{fig:corr1}, the correlations $\zeta_{22}(r)$ and
$\zeta_{33}(r)$  (recall that $\zeta_{11}=\zeta_{33}$) averaged over
the realizations are shown as  empty symbols. These measurements are
compared to the asymptotic scaling~(\ref{eq:ceig2}) and to the
following analytic estimates
\begin{eqnarray}
\bar{\zeta}_{11}(r) &=& \frac{1}{9}\,\psi(r)
+\left(\frac{87}{270}-\frac{9}{10\pi}\right)\psi^2(r) \nonumber \\
\bar{\zeta}_{22}(r) &=& \frac{1}{9}\,\psi(r)+\frac{1}{45}\,\psi^3(r) \;,
\label{eq:ceig3}
\end{eqnarray}
which are designed to asymptote to the variances given in
Eq.~(\ref{eq:vareig}). The density correlation function $\psi$
measured from the simulations is used for the evaluation of those
theoretical expressions. Figure~\ref{fig:corr1} clearly demonstrates
that, while the linear approximation $\psi/9$ is  in excellent
agreement with the measurements in the asymptotic regime, it deviates
at  least 10 percent at small distance, $r\lsim 3\hmpc$. By contrast,
$\bar{\zeta}_{ii}(r)$ as defined above achieves a fractional error no
larger than 2 percent for separations less than $\sim 30\hmpc$.

These results are readily extended to the cross correlations 
$\zeta_{ij}$, $i\ne j$. Proceeding in a similar way, $\zeta_{ij}(r)$ 
can be rearranged as
\begin{equation}
\zeta_{ij}(r)=\frac{1}{9}\,\psi(r)+\frac{1}{10}\la x_i\ra 
\la y_j\ra\,\psi^2(r)
\end{equation}
to second order in $\psi(r)$. Estimating the cross correlations at
zero lag $\zeta_{ij}(0)=\la x_i x_j\ra$ from the theoretical
probability distributions $P(x_i,x_j)$ proves difficult. A numerical 
integration  gives the following hypothesized rational forms
\begin{equation}
\la x_1 x_2\ra=\la x_2 x_3\ra=\frac{1}{10},~~
\la x_1 x_3\ra=\frac{27-6\pi}{30\pi}\;,
\end{equation}
for which the constraint $\la(x_1+x_2+x_3)^2\ra=1$ is satisfied. This
motivates the interpolation formulae
\begin{eqnarray}
\bar{\zeta}_{13}(r) &=& \frac{1}{9}\,\psi(r)
-\left(\frac{14}{45}-\frac{9}{10\pi}\right)\psi^2(r) \nonumber \\
\bar{\zeta}_{12}(r) &=& \bar{\zeta}_{23}(r) = \frac{1}{9}\,\psi(r)
-\frac{1}{90}\,\psi^3(r) \;,
\label{eq:ceig4}
\end{eqnarray}
which match reasonably well the large- and small-scale behavior of
$\zeta_{ij}$ (Fig.~\ref{fig:corr1}). Finite grid size effects may
be responsible for the slight offset (roughly 2 percent) of the
cross correlation $\zeta_{13}$ relative to the theoretical prediction. 

Numerical investigations indicate that dark matter haloes do not form
randomly in the initial conditions, but rather preferentially close to
the peaks of the density field~\cite{haloloc}.
To assess the extent to which biasing affects our results, we adopt
in a first approximation the usual critical density criterion issued
from the spherical infall model.  As first recognized
in~\cite{Kaiser1984}, the correlation function of regions lying above
a certain density threshold $\nu$ is enhanced relative to that of the 
density correlation $\psi(r)$. Likewise, the correlations  of shear
eigenvalues restricted to regions with density larger than a given
threshold $\nu$ are also  amplified. On large scales and in the regime
$\nu\gg 1$, we find
\begin{equation}
\zeta_{ii}(r|>\nu)\approx
\left(\frac{1}{3}+\nu\la x_i|\nu\ra\right)^2\psi(r)\;,
\label{eq:amp1}
\end{equation}
where the conditional average eigenvalue $\la x_i|\nu\ra$,
\begin{equation}
\la x_i|\nu\ra=\la x_i\ra+\frac{\nu}{3}\;,
\end{equation}
depends linearly on the peak height $\nu$. This should be compared to 
the correlation function of thresholded regions, which is 
$\xi(r|>\nu)\approx\nu^4\psi(r)$ in the same limit~\cite{Kaiser1984}.
Hence, $\zeta_{ii}(r|>\nu)$ also exhibits the usual linear amplification
factor $\nu^4$ of dense regions. It would be interesting to estimate 
the extent to which the large-scale bias varies when constraints are 
imposed on the shear eigenvalues. This calculation is postponed to
a future work.

\begin{figure}
\center \resizebox{0.45\textwidth}{!}{\includegraphics{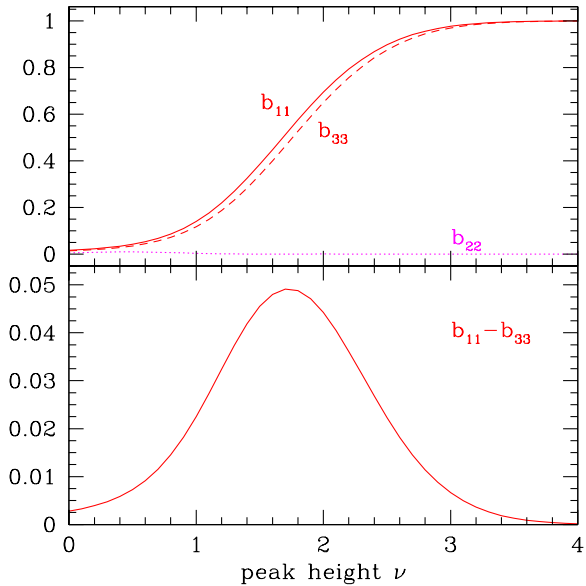}}
\caption{Correction factors $b_{ii}(\Delta,\nu)$  (see text) as a
function of the threshold height $\nu$ when the alignment of principal
axes is restricted to those regions where all shear eigenvalues are
positive. The bottom panel shows the difference $b_{11}-b_{33}$, which
demonstrates that the alignment between major axes is slightly stronger 
than the correlation of minor axes. The correlation  of the intermediate 
axis is strongly suppressed over the whole range of peak height.} 
\label{fig:etan}
\end{figure}

\subsection{Principal axes}

We now turn to the correlation of the shear principal axes. Since it
is  computationally expensive to measure such a correlation from
numerical realizations (direct summation must be employed), we will
only present an analytic estimate which is valid on large scales. We
also discuss its dependence on the peak height. Note that similar
calculations for the correlation of angular momentum
$L_i\propto\epsi_{ijk}\xi_{jl}I_{lk}$, where $I_{lk}$ is the inertia
tensor of some Lagrangian region, can be found in
~\cite{LeePen2001,Crittendenetal2001} for instance.

Let $\nvh_i$ designate the unit vector in the direction of the major, 
intermediate, or minor axis of the shear. It is conventional to use 
\begin{equation}
\eta_{ij}(r) = \la|\nvh_i(\vq_1).\nvh_j(\vq_2)|^2\ra-\frac{1}{3}
\end{equation}
as a measure of the alignment between principal axes
~\cite{Penetal2001,LeePen2007}, so that $\eta_{ij}(r)=0$ in the 
absence of any correlation. 

\begin{figure*}
\resizebox{0.45\textwidth}{!}{\includegraphics{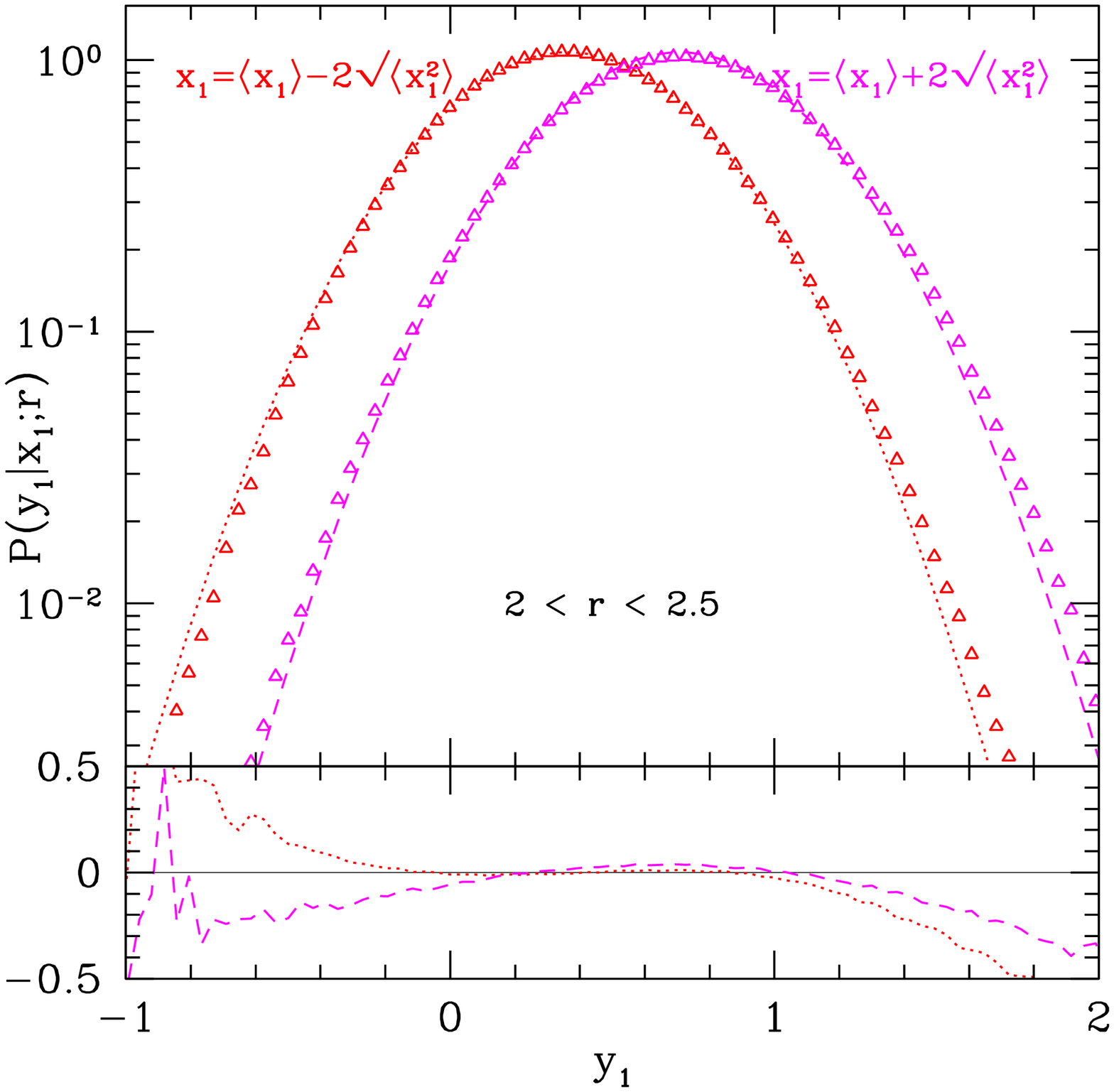}}
\resizebox{0.45\textwidth}{!}{\includegraphics{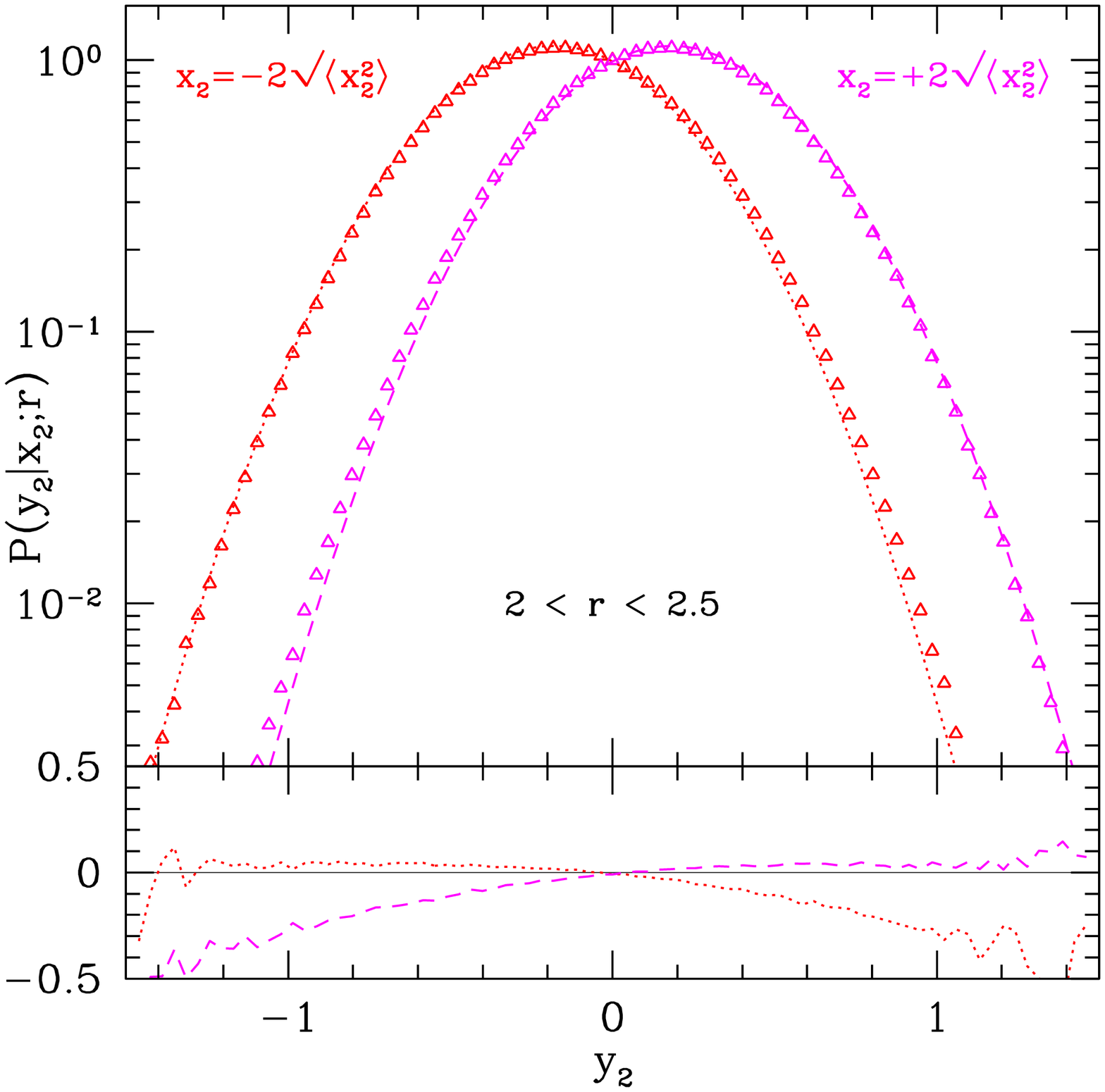}}
\caption{A comparison between the conditional 2-point probability 
distribution $P(y_i|x_i;r)$ measured from realizations of Gaussian random 
fields and the Gaussian approximation equation~(\ref{eq:cpdf1}). 
Results are presented for the largest and intermediate eigenvalues only 
(left and right panels respectively). Symbols show the measurements of 
the conditional probability for a separation $2<r<2.5\hmpc$ and for 
$x_i=\la x_i\ra\pm 2\tilde{\sigma}_i$. The correlation $\zeta_{ii}(r)$ used 
in the analytic estimate shown as the dashed curve is directly calculated 
from the random realizations of the shear field. The bottom panels show 
the fractional error.}
\label{fig:pdf1}
\end{figure*}

We will only consider the correlations $\eta_{ii}$ since the
calculation  of $\eta_{ij}$ proceeds along similar lines. We
parametrise the rotation matrix $\vrr$ in terms  of the Euler angles
$0\leq \varphi,\psi\leq 2\pi$ and $0\leq\vartheta\leq\pi$. We adopt
the  XYX, YZY, and ZYZ convention when working out the correlation  of
major, intermediate and minor axes, respectively, so that
$\cos\vartheta$ always is the cosine  of the angle between the
considered axes.  The average is performed over the independent
components of the shear  tensor,
\begin{equation}
\eta_{ii}(r) = \int\!\!\dd\xi_1\dd\xi_2\,
\left(\cos^2\!\vartheta-\frac{1}{3}\right)\,
P\left(\xi_1,\xi_2;r\right)\;,
\end{equation} 
where, in the limit $r\gg 1$, the 2-point probability distribution
$P(\xi_1,\xi_2;\vr)$ reduces to Eq.~(\ref{eq:problargescale}). The
quadratic form $Q_1(\Lambda_1,\vrr\Lambda_2\vrr^\top)$ can be
expanded in terms of the Wigner D-functions  ${\cal
D}^l_{_{m_1,m_2}}$, $l$ being the index of the representation. These
3D  harmonics generate irreducible representations of the
three-dimensional rotation group and, therefore, form  a complete
orthogonal set of functions defined on SO(3) itself. Invariance under
reflections  implies that the quadrupole rotation matrices with
$m=0,\pm 2$  appear in the harmonic decomposition of the quadratic
form $Q_1$ as follows~:
\begin{eqnarray}
\lefteqn{Q_1(\Lambda_1,\vrr\Lambda_2\vrr^\top)=\frac{1}{2}\,
\tr\Lambda_1\tr\Lambda_2+\frac{15\epsi_-}{8}\left[{\cal D}^2_{_{0,0}}
\right.}\nonumber  \\ && - \left.\sqrt{\frac{3}{2}}\epsi_1\left({\cal
D}^2_{_{0,-2}} +{\cal D}^2_{_{0,2}}\right)-\sqrt{\frac{3}{2}}\epsi_2
\left({\cal D}^2_{_{-2,0}}+{\cal D}^2_{_{2,0}}\right)\right.\nonumber
\\  && + \left.\frac{3}{2}\,\epsi_1\epsi_2\left({\cal D}^2_{_{-2,-2}}
+{\cal D}^2_{_{2,2}}+{\cal D}^2_{_{-2,2}}+{\cal D}^2_{_{2,-2}} \right)
\right]\;.
\label{eq:trace1}
\end{eqnarray}
The explicit form of these harmonics in the ZYZ representation is given 
in Table~\ref{table:table1}.
It is worth emphasizing that, apart from the ``traceful'' contribution 
$1/2\,\tr\Lambda_1\tr\Lambda_2$, 
$Q_1(\Lambda_1,\vrr\Lambda_2\vrr^\top)$ depends on the three ``shape''
parameters $\epsi_-$, $\epsi_1$, and $\epsi_2$ solely, because  points
on SO(3) truly have only 3 degrees of freedom. These  parameters
are given by  $\epsi_-=(1/3)\,(x_{13}+x_{23})(y_{13}+y_{23})$, 
$\epsi_1=(x_1-x_2)/(x_{13}+x_{23})$, $\epsi_2=(y_1-y_2)/(y_{13}+y_{23})$, 
where $x_{ij}=x_i-x_j$ and $y_{ij}=y_i-y_j$.  The alignment is simply
$\eta_{ij}=2/3\,\la {\cal D}^2_{_{0,0}}\ra$ with our parametrization of
the rotation matrix. The addition of angular momentum then yields
\begin{equation}
\frac{1}{4}\int_{\rm SO(3)}\!\!\!\!\!\!\dd\vrr\,
\left(\cos^2\vartheta-\frac{1}{3}\right)\,
Q_1(\Lambda_1,\vrr\Lambda_2\vrr^\top)=\frac{\epsi_-}{16}\;.
\end{equation}
after averaging over the angular variables. Lastly, the integral over 
the volume measure $\dd^3x\dd^3y$ is easily performed in the coordinate 
system $(\nu,e,p)$, where 
\begin{equation}
\epsi_-=
\left\{\begin{array}{c}\frac{1}{3}\nu_1\nu_2\left(3e_1-p_1\right)
\left(3e_2-p_2\right)~~~\mbox{minor axis} \\ \frac{1}{3}\nu_1\nu_2
\left(3e_1+p_1\right)\left(3e_2+p_2\right)~~~\mbox{major axis} 
\end{array}\right.\;.
\end{equation}
Let us choose $\nu_1=\nu_2=\nu$ for illustration purposes, and perform 
the integration over the  domain defined by $e_i\geq 0$ and $|p_i|\leq
e_i$ ($i=1,2$). We find  that, in the large-scale limit $r\gg 1$, the  
alignment $\eta_{ii}(r|>\nu)$ of thresholded regions evaluates to
\begin{equation}
\eta_{ii}(r|>\nu)=\frac{27}{20\pi}\,\psi(r)\approx 0.43\,\psi(r)\;,
\label{eq:align1}
\end{equation}
regardless of the peak height and the axis under consideration (major
and minor). This  is unsurprising given the invariance of the integral 
over the asymmetry parameters under the reflection $p_i\rightarrow
-p_i$. For the  intermediate axis, a similar calculation yields
$\epsi_-=(4/3)\nu_1\nu_2 p_1 p_2$. This implies $\eta_{22}(r|>\nu)=0$ 
at leading order since the 1-point probability $P(e,p|\nu)$, 
\begin{equation}
P(e,p|\nu)=\frac{1125}{\sqrt{10\pi}}\,\nu^5 e\left(e^2-p^2\right)
e^{-\frac{5}{2}\nu^2\left(3 e^2+p^2\right)}\;,
\label{eq:pepv}
\end{equation}
is symmetric about $p=0$. 

\begin{table*}
\caption{Quadrupole Wigner D-functions  ${\cal
D}^2_{_{m_1,m_2}}\!\left(\varphi,\vartheta,\psi\right)$ in the ZYZ
representation. Harmonics with $m_1,m_2=\pm 1$ are not shown since they
are not needed for the present analysis.}
\begin{center}
\begin{tabular}{llll} \hline\hline
&  $m_2=-2$ & $m_2=0$ & $m_2=2$ \\ \hline $m_1=-2$ &
$\frac{1}{4}\left(1+\cos\vartheta\right)^2 e^{2i\varphi+2i\psi}$ &
$\sqrt{\frac{3}{8}}\sin^2\!\vartheta\,  e^{2i\varphi}$ &
$\frac{1}{4}\left(1-\cos\vartheta\right)^2 e^{2i\varphi-2i\psi}$ \\
$m_1=0$ & $\sqrt{\frac{3}{8}}\sin^2\!\vartheta\,  e^{2i\psi}$ &
$\frac{1}{2}\left(3\cos^2\!\vartheta-1\right)$ &
$\sqrt{\frac{3}{8}}\sin^2\!\vartheta\,e^{-2i\psi}$ \\ $m_1=2$ &
$\frac{1}{4}\left(1-\cos\vartheta\right)^2 e^{-2i\varphi+2i\psi}$  &
$\sqrt{\frac{3}{8}}\sin^2\!\vartheta\, e^{-2i\varphi}$ & $\frac{1}{4}
\left(1+\cos\vartheta\right)^2 e^{-2i\varphi-2i\psi}$ \\ \hline\hline
\end{tabular}
\end{center}
\label{table:table1}
\end{table*}

Unlike the conditional correlation $\zeta_{ii}(r|>\nu)$ of shear 
eigenvalues,  the alignment $\eta_{ii}(r|>\nu)$ between the shear
principal axes is insensitive to the peak height. However, restricting
the domain of   integration to the region where all shear eigenvalues
are positive,  for instance, can introduce a dependence on the threshold
height.  Such a constraint naturally arises in models of structure
formation  to characterize the Lagrangian regions which collapse into
dark matter haloes.  The domain where  the lowest eigenvalue is positive
corresponds to the  interior of the  triangle bounded by
$(e,p)=(0,0)$,  $(\frac{1}{4},-\frac{1}{4})$, and
$(\frac{1}{2},\frac{1}{2})$.  The conditional correlation can be
conveniently expressed as $b_{ii}(\Delta,\nu)\eta(r|>\nu)$, where
$\eta(r|>\nu)=27/20\pi\,\psi(r)$ and $b_{ii}(\Delta,\nu)$ is a 
correction factor resulting from the restriction to the triangular 
domain $\lambda_3>0$. After some manipulation, we find
\begin{eqnarray}
b_{11}(\Delta,\nu) &\approx&
\left\{1-{\rm Erf}\!\left(\frac{\nu}{\sqrt{2}}\right)^{-1}\!
\left[\sqrt{\frac{2}{\pi}}\frac{5\nu}{27}\,e^{-9\nu^2/8} 
\right.\right. \nonumber  \\ 
\lefteqn{\left.\left.+\frac{\sqrt{60}}{27}
{\rm Erf}\!\left(\sqrt{\frac{15}{2}}\frac{\nu}{2}\right)e^{-9\nu^2/8}
+\frac{34}{81}{\rm Erfc}\!\left(\frac{3\nu}{2\sqrt{2}}\right) \right.
\right. } \nonumber \\ \lefteqn{\left.\left. +\frac{5\sqrt{6}}{27}
{\rm Erfc}\!\left(\sqrt{3}\nu\right) \right]\right\}^2} \\
b_{22}(\Delta,\nu) &\approx&
\left\{\left[\sqrt{\frac{2}{\pi}}\frac{5\nu}{27}
-\frac{\sqrt{60}}{27}{\rm Erf}\!\left(\sqrt{\frac{15}{2}}
\frac{\nu}{2}\right) \right] \right. \nonumber \\
\lefteqn{\left. \times e^{-9\nu^2/8} +\frac{34}{81}
{\rm Erfc}\!\left(\frac{3\nu}{2\sqrt{2}}\right)-\frac{1}{\sqrt{6}}
{\rm Erfc}\!\left(\sqrt{3}\nu\right)\right\}^2} \nonumber \\
\lefteqn{{\times \rm Erf}\!\left(\frac{\nu}{\sqrt{2}}\right)^{-2}} \\ 
b_{33}(\Delta,\nu) &\approx&
\left\{1-{\rm Erf}\!\left(\frac{\nu}{\sqrt{2}}\right)^{-1}\!
\left[\sqrt{\frac{2}{\pi}}\frac{10\nu}{27}\,e^{-9\nu^2/8} 
\right.\right. \nonumber  \\ \lefteqn{\left.\left.+\frac{68}{81}
{\rm Erfc}\!\left(\frac{3\nu}{2\sqrt{2}}\right)+\frac{\sqrt{6}}{54}
{\rm Erfc}\!\left(\sqrt{3}\nu\right) \right]\right\}^2\;.}
\label{eq:align2}
\end{eqnarray}
These expressions are plotted in the upper panel of
Fig.~\ref{fig:etan} as a function of the threshold height. Clearly,
the alignment between major and minor axes is strongly suppressed when
$\nu\lsim 2$.  The correction factors  $b_{ii}(\Delta,\nu)$ are always
less than one, to which they asymptote in the limit of large threshold
$\nu$.  Furthermore, in the range $\nu\sim 1-3$, we have $b_{11}\bsim
b_{33}$  so that the correlation between major axes is stronger by a
few per  cent. Shown in the lower panel of Fig.~\ref{fig:etan} is the
difference $b_{11}-b_{33}$. It is maximal when the integral of the
probability distribution $P(e,p|\nu)$  over the triangle is about 
one-half. This occurs when the mean  ellipticity $\la
e|\nu\ra=3/(\sqrt{10\pi}\nu)$ roughly is $1/4-1/2$,   i.e. when
$\nu\approx 2$. Restricting the integration domain to $\lambda_3>0$
thus induces a nonzero, albeit small, correlation between
intermediate axes. $\eta_{22}(r|>\nu)$ does not exceed $\sim
0.01\psi$,  a value reached when $\nu\sim 0.5$. We emphasize that the
alignment between the principal axes of the shear field is in all
cases proportional to the density correlation on large scales. This
is consistent with the findings of
~\cite{Crittendenetal2001,HirataSeljak2004}.

We have shown that the initial alignment of the shear principal axes
is significant only for those regions which collapse into haloes of
mass $M\bsim M_\star$.  We will, nevertheless, not extend the discussion 
to the statistics of dark  matter halo and galaxy shapes  as  it is
unclear to which extent the  correlations detected in  N-body
simulations of CDM cosmologies~\cite{CroftMetzler2000}, or in galaxy
surveys ~\cite{Heavensetal2000}, reflect the large-scale alignment of
the initial shear. Nonlinear effects, such as anisotropic accretion or
relaxation following collapse, could plausibly enhance or erase the
large-scale coherence of the primordial tidal
field~\cite{West1989,HuiZhang2002}. Indeed, even in the artificial 
case where all dark matter haloes are perfectly aligned would the
alignment and clustering of galaxies be negligibly affected 
~\cite{align}. We conclude by noticing that primordial
non-Gaussianities characterized by a local mapping of the form
$\phi_{\rm NG}=\phi-f_{\rm NL}(\phi^2-\la\phi^2\ra)$
~\cite{Matarreseetal2000,KomatsuSpergel2001,Maldacena2003} would not
affect the correlation of principal axes.

\section{A Gaussian approximation}
\label{sec:gaussian}

\subsection{Shear eigenvalues}

Explicit expressions for  the 1-point probability distribution
$P(x_i)$ of the individual shear eigenvalues $x_i$ can be found in
~\cite{Doroshkevich1970,LeeShandarin1998}. It is worth
noticing that , although the variables $x_i(\vq)$ are not Gaussian
random fields, their 1-point probability distributions are very
close to Gaussian. More precisely, $P(x_2)$ is indeed (fortuitously)
exactly Gaussian, whereas the  probability densities  $P(x_1)$ and
$P(x_3)$ show a small positive skewness,
\begin{equation}
{\rm skewness}=\frac{3^{3/2}\left(54-17\pi\right)}
{\left(13\pi-27\right)^{3/2}}\approx 0.060\;,
\end{equation}
which reflects the fact that the large and small tail of these 
distributions, respectively, is slightly more pronounced. 

\begin{figure}
\center \resizebox{0.45\textwidth}{!}{\includegraphics{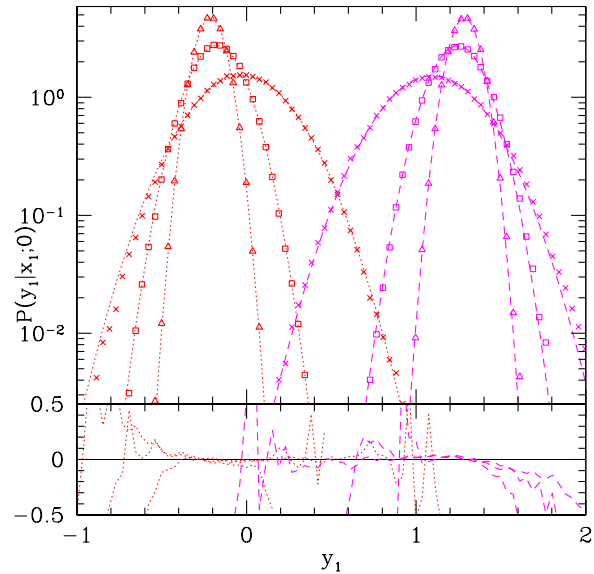}}
\caption{The conditional probability density $P(y_1|x_1;0)$ evaluated
at a single Lagrangian position but at two different smoothing lengths
$R_1$ and $R_2$. The triangles, squares, and crosses show
$P(y_1|x_1;0)$  measured from random  realizations of the potential
smoothed on scale  $R_2=2.5$, 3, and 5$\hmpc$, respectively, while
keeping $R_1=2\hmpc$ fixed. The dotted and dashed curves indicate the
Gaussian approximation when  $x_1=\la x_1\ra+2\tilde{\sigma}_1$ and
$\la x_1\ra-2\tilde{\sigma}_1$, respectively. The bottom panel shows
the fractional error.}
\label{fig:pdf3}
\end{figure}

To assess the extent to which 2-point statistics of the shear
eigenvalues deviate from Gaussianity, we  have measured the
conditional, 2-point probability density  $P(y_i|x_i;r)$ from random
realizations of the potential field. Results are presented in
Fig.~\ref{fig:pdf1} (symbols) for a separation in the range
$2<r<2.5\hmpc$, and  for two particular values of  $x_i$~: $x_i=\la
x_i\ra\pm 2\tilde{\sigma}_i$ where $\tilde{\sigma}_i^2\equiv\la
x_i^2\ra$ in what follows.  These measurements are compared to the
following Gaussian conditional  distribution~:
\begin{eqnarray}
P(y_i|x_i;r) &=& \frac{\tilde{\sigma}_i}
{\sqrt{2\pi\left(\tilde{\sigma}_i^4-\zeta_{ii}^2\right)}} \\ && 
\times\,\exp\left[-\frac{\tilde{\sigma}_i^2 (\tilde{x}_i^2+\tilde{y}_i^2)
-2\zeta_{ii} \tilde{x}_i \tilde{y}_i}
{2\left(\tilde{\sigma}_i^4-\zeta_{ii}^2\right)}+\frac{\tilde{x}_i^2}
{2\tilde{\sigma}_i^2}\right] \nonumber\;,
\label{eq:cpdf1}
\end{eqnarray}
where, for shorthand convenience, we have introduced the variable
$\tilde{x}_i\equiv x_i-\la x_i\ra$.  This conditional probability is
shown as the dashed curves. Figure~\ref{fig:pdf1} nicely demonstrates
the validity of the Gaussian approximation down to scales  of the
order of the grid resolution (about $1\hmpc$). We also note that, in
the distribution $P(y_1|x_1;r)$, the skewness increases by $\sim 50$
percent to reach $\approx 0.089$ at separation $r\sim 2\hmpc$
whereas  in  $P(y_2|x_2;r)$, the skewness is approximately 0.016 at
the same   distance.

Thus far, we have restricted the comparison to the case in which the
shear at Lagrangian positions $\vq_1$ and $\vq_2$ is smoothed at the
same length $R_S$. To further assess the validity of the Gaussian
approximation in the limit $\zeta_{ii}\approx \tilde{\sigma}_i^2$, we
take advantage of the fact that the joint distribution of the shear
eigenvalues, evaluated at a single Lagrangian position $\vq_1=\vq_2$
and two different smoothing lengths $R_1$ and $R_2$, also has the
functional form of Eq.~(\ref{eq:2pdfc}), with $\psi=\gamma$
~\cite{Desjacques2007}. Therefore, despite the limitation arising from
the finite grid spacing, we can nevertheless probe the strongly
correlated regime by studying the conditional probability density
$P(y_i|x_i;0)$ in the limit $R_1\approx R_2$. For illustration, let us
consider the largest eigenvalue. We set $R_1=2\hmpc$ and take
$R_2=2.5$, 3 and 5$\hmpc$.  The cross correlation $\zeta_{11}$ at two
different smoothing scales,  computed following the procedure outlined
in Sec.~\ref{sec:correl}, yields $\tilde{\sigma}_1^2=0.143$, 0.136, and
0.108~\cite{zetalimit}, respectively. These values are used for the
evaluation of the conditional density~(\ref{eq:cpdf1}). As seen from
Fig.~\ref{fig:pdf3}, there is, as before, a remarkably good agreement
between $P(y_1|x_1;0)$ measured from random realizations of the
potential $\Phi$ and the conditional bivariate Gaussian
equation~(\ref{eq:cpdf1}), except for the very tails of the
distributions. This strongly suggests that Gaussian statistics are an
accurate approximation to the statistics of shear eigenvalues at all
separations and smoothing scales.

\begin{figure}
\center \resizebox{0.45\textwidth}{!}{\includegraphics{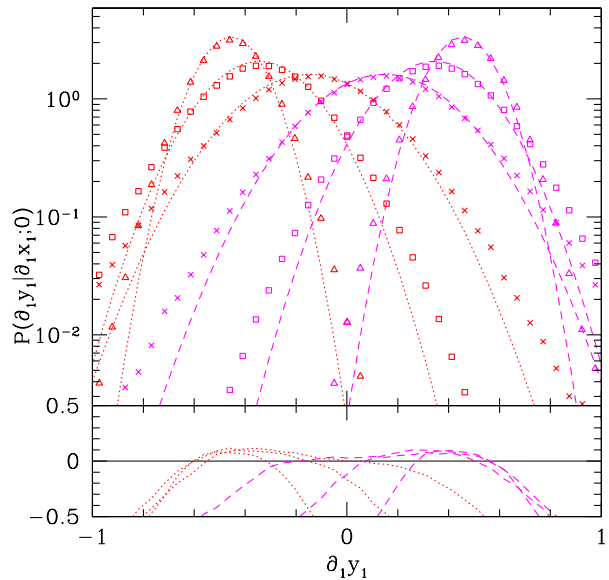}}
\caption{Same as Fig.~\ref{fig:pdf3} but showing the conditional
probability  density
$P(\partial_1\tilde{y}_1|\partial_1\tilde{x}_1;0)$. Deviations  from
the bivariate Gaussian are more pronounced here than in the
conditional distribution $P(y_1|x_1;0)$.}
\label{fig:pdf4}
\end{figure}

\subsection{Gradients of the shear eigenvalues}

The derivatives of any Gaussian random field $X(\vq)$ with respect to
the coordinate $\vq$ also are Gaussian random fields since the  
differential operators $\partial/\partial q_i$ are linear. To 
ascertain how much the spatial derivatives of the shear eigenvalues
deviate from Gaussianity, we have examined a number of  correlation 
functions and conditional 2-point statistics, focusing on the  first 
derivatives $\partial_i x_1$ of the largest eigenvalue. This isotropic
vector field has a covariance tensor of the form 
\begin{equation}
\la\partial_i x_1\partial_j y_1\ra=\Sigma_1(r)\rh_i\rh_j
+\Sigma_2(r)\delta_{ij}\;.
\end{equation}  
When $\zeta_{11}$ is known analytically, exact expressions can be derived 
fairly  easily using the relation
$\la\partial_i x_1\partial_j y_1\ra=-\partial_i\partial_j\zeta_{11}(r)$. 
For instance, the approximation~(\ref{eq:ceig3}) gives
\begin{eqnarray}
\lefteqn{\la\partial_i x_1\partial_j y_1\ra\approx 
\left(\frac{1}{9}+c\,\psi^2\right)\delta_{ij}} && \\ && 
+ \left[2\,c\,\psi^{'2}+\left(\frac{1}{9}+2\,c\,\psi\right)
\left(-\frac{\psi'}{9}+\psi^{''}\right)\right]\rh_i\rh_j \nonumber \;,
\label{eq:cdeig1}
\end{eqnarray}
where $c=87/270-9/10\pi$ and a prime denotes a derivative with respect
to $r$. Let us introduce the dimensionless variable 
$\partial_i\tilde{x}_1\equiv(\sigma_0/\sigma_1)\partial_i x_1$, where 
the $\sigma_j$ generally are the spectral moments of  the density field
~\cite{Bardeenetal1986},
\begin{equation}
\sigma_j^2 \equiv \int_0^\infty\!\!\dd\ln k\,k^{2j}\,\Delta_\delta^2(k)\;.
\label{eq:mspec}
\end{equation}
Using Eq.~(\ref{eq:cdeig1}) as an analytic estimate of the correlation 
$\la\partial_i\tilde{x}_1\partial_i\tilde{y}_1\ra$, we find a variance
\begin{equation}
\la\partial_i\tilde{x}_1^2\ra=
\frac{1}{3}\left(\frac{34}{45}-\frac{9}{5\pi}\right)
\approx 6.087 \times 10^{-2}\;,
\end{equation}    
somewhat 10 per cent smaller than the value of $\approx 6.60 \times
10^{-2}$ measured from the random realizations with $R_S$ in the range
$2-5\hmpc$.  The agreement can be improved by adding higher order
terms in the  truncated expansion equation~(\ref{eq:ceig3}).  

The conditional probability
$P(\partial_1\tilde{y}_1|\partial_1\tilde{x}_1;0)$  is shown in
Fig.~\ref{fig:pdf4} for three different smoothing lengths $R_1\approx
R_2$ (as in Fig~\ref{fig:pdf3}). Cross correlations coefficients  and
variances are computed from the random realizations. Clearly, although
deviations from Gaussianity are more pronounced than in the
conditional 2-point statistics of the eigenvalues discussed above, 
the agreement is still reasonable. However, we have found that it
worsens significantly for the second derivative. In spite of this
limitation, it would be valuable to assess whether the Gaussian 
approximation provides a reliable description of the statistics of 
extrema of the shear eigenvalues, since the latter play a particular
role in nonspherical collapse models~\cite{shearfield}.

\section{Conclusion}
\label{sec:conclusion}

We have explored the statistical correlation that arises in Gaussian
initial conditions between the properties of the linear tidal shear
$\partial_i\partial_j\Phi$ at two distinct positions, thereby
extending the work of~\cite{Doroshkevich1970,Bardeenetal1986,
LeeShandarin1998,CatelanPorciani2001,Desjacques2007}. In
Sec.~\ref{sec:2pdf}, using asymptotic expansions, we derived
exact  closed form expressions for the joint distribution  of shear
components and shear eigenvalues as a function of the Lagrangian
separation.  These results were applied to study the large-distance
asymptotics of  the correlation function of the shear eigenvalues and
the shear  principal axes. In Sec.~\ref{sec:correl}, we presented
interpolation formulae that accurately match the large- and
small-scale behavior  of the correlation of shear eigenvalues measured
from random realizations of the gravitational potential. We also found
that the alignment of the shear principal axes of thresholded regions
is insensitive to  the threshold height. However, restricting the
correlation to  regions where all three eigenvalues are positive,
introduces a  dependence on the threshold density, which manifests
itself as a strong suppression of the alignment for peak height less
than $\nu\sim 1$. We emphasize that all these  correlations are
proportional to the density correlation on large  scale.

In Sec.~\ref{sec:gaussian}, we showed  that the 2-point
statistics of the shear eigenvalues closely follow the Gaussian
statistics regardless of the separation and the smoothing
length. Although we have not formally established that Gaussian
multivariates comply with measurements of the $n$-point  distributions
of shear eigenvalues, we speculate that the Gaussian approximation
also holds for these multipoint distributions. Under this assumption,
it should be fairly straightforward to apply the techniques and
results obtained for Gaussian density fields to the shear eigenvalues.

Gaussian statistics provides also a reasonable  description of a
number of conditional probability densities involving first
derivatives of the shear eigenvalues. Note, however, that the
agreement worsens noticeably  for the second derivatives. This caveat
notwithstanding, a Gaussian approximation should be adequate to
understand, at least qualitatively,  the clustering of extrema of the
shear eigenvalues for instance. The mathematical framework laid down
by~\cite{Bardeenetal1986} appears well suited for such a study.

Our results can also be applied to the description of large-scale
structures using the cosmic web approach based on the ellipsoidal
collapse~\cite{BondMyers1996,Bondetal1996}.  In light of our analysis,
the conditional multivariate Gaussian describing the joint
distribution of the density, displacement field, and shear could
easily be written down. As recognized in~\cite{Bondetal1996}, these
statistics will prove useful for quantifying the properties of the
mildly nonlinear fluctuations, which evolve into the network of
clusters, filaments, and walls observed in the recent 2dF and SDSS
galaxy surveys ~\cite{galaxysurvey}.  Constraints on the tidal shear
could also be included in  topological measures such as Minkowski
functionals, in an attempt to study the effect of nonspherical infall
on the morphology of the  primeval large-scale structures.

\acknowledgements

We acknowledge support by the Swiss National Foundation under 
Contract No.~200021-116696/1.

\appendix

\section{Averaging over the relative orientations}
\label{app:orientation}

The integral~(\ref{eq:intso3}) over the relative orientation of the
principal axis frames can be expressed as a hypergeometric function
with the symmetric 3$\times$3 matrices $\beta\Lambda_1$ and
$\Lambda_2$ as argument,
\begin{equation}
_0F_0^{(3)}\!\left(\beta\Lambda_1,\Lambda_2 \right)=
\sum_{k=0}^\infty\frac{\beta^k}{k!}\sum_{\kappa}
\frac{C_\kappa(\Lambda_1)C_\kappa(\Lambda_2)}{C_\kappa(\vii)}\;.
\label{eq:intso3form2}
\end{equation}
Here $\sum_\kappa$ designates summation over all partitions
$\kappa\vdash k$ of $k$, namely, over the ordered sequence of integer
$(k_1,k_2,\dots,k_n)$ such that $ k_1\geq k_2\geq\dots\geq k_n\geq 0$
and $\sum k_i=k$. $C_\kappa(\vxx)$ are the zonal polynomials of the
matrix $\vxx$. They satisfy the relation $(\tr\vxx)^k=\sum_\kappa
C_\kappa(\vxx)$. We emphasize that, despite the use of matrix notation
here, $C_\kappa(\vxx)$ is a function of the eigenvalues of $\vxx$
solely and could thus be written as $C_\kappa(x)$ for example.

The zonal polynomial can be expressed in terms of the monomial
symmetric functions  $m_\kappa(x)$. When $k=2$ for instance, there are
two zonal polynomials  corresponding to the partitions $(2)$ and
$(1,1)$ of 2, $C_{(2)}=m_{(2)}(x)+2/3\, m_{(1,1)}(x)$ and
$C_{(1,1)}=4/3\, m_{(1,1)}$, where
\begin{eqnarray}
m_{(2)}(x) &=& x_1^2+x_2^2+x_3^2 \nonumber \\  m_{(1,1)}(x) &=& x_1
x_2+ x_1 x_3+ x_2 x_3 \;.
\label{eq:esym}
\end{eqnarray}
The value of these zonal polynomials at $\vii$ is $C_{(2)}(\vii)=5$ and
$C_{(1,1)}(\vii)=4$.  There is a recurrence relation between the
coefficients of $m_\kappa(x)$  that determines $C_\kappa(\vxx)$
uniquely once the coefficient of $m_{(k)}$  is
given~\cite{Muirhead1982,fastcomp}.  Note also that the functions
$m_\kappa(x)$ can be written in terms of the traces of power of $\vxx$, 
$\tr\vxx^k$ with $k=0,1,\dots$. We refer the reader to
~\cite{Muirhead1982} for further details.


\begin{thebibliography}{50}

\bibitem[\protect\citeauthoryear{}{}]{galaxysurvey}{
M.~Colless, et al., 
ArXiv Astrophysics eprint:astro-ph/0306581 (2003);
J.K.~Adelman-McCarthy, et al., 
ArXiv Astrophysics eprint:astro-ph/0707.3413 (2007).
\label{galaxysurvey}}

\bibitem[\protect\citeauthoryear{}{}]{simulations}{
C.~Park, \MNRAS, {\bf 242}, 59 (1990);
A.~Nusser, A.~Dekel, \ApJ, {\bf 362}, 14 (1990);
D.H.~Weinberg, J.E.~Gunn, \MNRAS, {\bf 247}, 260 (1990);
E.~Bertschinger, J.M.~Gelb, Comp.~Phys.~, {\bf 5}, 164 (1991);
R.~Cen, J.P.~Ostriker, \ApJ, {\bf 417}, 415 (1993);
A.~Jenkins, et al., \ApJ, {\bf 499}, 20 (1998);
V.~Springel, et al., \Nat, {\bf 435}, 629 (2005);
S.~Gottl\"ober, G.~Yepes, \ApJ, {\bf 664}, 117 (2007).
\label{simulations}}

\bibitem[\protect\citeauthoryear{Gunn \& Gott}{1972}]{GunnGott1972}
J.E.~Gunn, J.R.~Gott III, \ApJ, {\bf 176}, 1 (1972).

\bibitem[\protect\citeauthoryear{}{}]{analytics}{
W.H.~Press, P.~Schechter, \ApJ, {\bf 187}, 425 (1974);
J.R.~Bond, S.~Cole, G.~Efstathiou, N.~Kaiser, \ApJ, {\bf 379}, 440 (1991);
C.~Lacey, S.~Cole, \MNRAS, {\bf 262}, 627 (1993);
H.J.~Mo, S.D.M.~White, \MNRAS, {\bf 282}, 347 (1996);
R.K.~Sheth, G.~Tormen, \MNRAS, {\bf 308}, 119 (1999).
\label{analytics}}

\bibitem[\protect\citeauthoryear{Peebles}{1980}]{Peebles1980}
P.J.E~Peebles, The Large-Scale Structure of the Universe (Princeton
University Press, 1980).

\bibitem[\protect\citeauthoryear{}{}]{topologics}{
J.R.III~Gott, A.L.~Melott, M.~Dickinson, \ApJ, {\bf 306}, 341 (1986);
K.R.~Mecke, T.~Buchert, H.~Wagner, \AA, {\bf 288}, 697 (1994);
T.~Matusbara, \ApJ, {\bf 434}, L43 (1994);
J.~Schmalzing, T.~Buchert, \ApJ, {\bf 482}, L1 (1997);
M.~Kerscher, et al., \MNRAS, {\bf 284}, 73 (1997);
M.~Kerscher, Lecture Notes in Physics, {\bf 554}, 36 (2000).
L.~Gleser, A.~Nusser, B.~Ciardi, V.~Desjacques, \MNRAS, {\bf 370},
1329 (2006);
J.R.III~Gott, et al., \ApJ, {\bf 675}, 16 (2008).
\label{topologics}}

\bibitem[\protect\citeauthoryear{}{}]{algorithms}{
V.~Sahni, B.S.~Sathyaprakash, S.F.~Shandarin, \ApJ, {\bf 495}, L5 (1998);
S.~Colombi, D.~Pogosyan, T.~Souradeep T., \PRL, {\bf 85}, 5515 (2000);
H.~Hanami, \MNRAS, 327, 721 (2001);
E.~Platen, R.~van de Weygaert, B.J.T~Jones, \MNRAS, {\bf 380}, 551 (2007);
M.A.~Arag\'on-Calvo, B.J.T.~Jones, R.~van de Weygaert, J.M.~van der Hulst, 
\AA, {\bf 474}, 315 (2007);
T.~Sousbie, C.~Pichon, S.~Colombi, D.~Novikov, D.~Pogosyan, \MNRAS,
{\bf 383}, 1655 (2008).
\label{algorithms}}

\bibitem[\protect\citeauthoryear{Bond, Kofman \& Pogosyan}{1996}]
{Bondetal1996} J.R.~Bond, L.~Kofman, D.~Pogosyan, \Nat, {\bf 380},
603 (1996).

\bibitem[\protect\citeauthoryear{}{}]{shearfield}{
Y.B.~Zeldovich, \AA, 5, 84 (1970);
V.I.~Arnold, S.F.~Shandarin, Y.B.~Zeldovich, Geophys. Astrophys. Fluid 
Dyn.~, {\bf 20}, 111 (1982);
S.D.M.~White, \ApJ, {\bf 286}, 38 (1984);
Y.~Hoffman, \ApJ, {\bf 308}, 493 (1986);
L.~Kofman, D.~Pogosyan, S.F.~Shandarin, A.L.~Melott, \ApJ, {\bf 393},
437 (1992);
R.~van de Weygaert, A.~Babul, \ApJ, {\bf 425}, L59 (1994);
E.~Bertschinger, B.~Jain, \ApJ, {\bf 431}, 486 (1994);
E.~Audit, J.-M.~Alimi, \AA, {\bf 315}, 11 (1996).
\label{shearfield}}

\bibitem[\protect\citeauthoryear{Sheth \& Tormen}{2002}]{ShethTormen2002}
R.K.~Sheth, G.~Tormen, \MNRAS, {\bf 329}, 61 (2002).

\bibitem[\protect\citeauthoryear{Binggeli}{1982}]{Binggeli1982}
B.~Binggeli, \AA, {\bf 107}, 338 (1982).

\bibitem[\protect\citeauthoryear{Argyres et al.}{1986}]{Argyresetal1986}
P.C.~Argyres, E.J.~Groth, P.J.E.~Peebles, M.F.~Struble, \AJ, {\bf 91},
471 (1986).

\bibitem[\protect\citeauthoryear{West et al.}{1989}]{Westetal1989}
M.J.~West, A.~Dekel, A.~Oemler, \ApJ, {\bf 336}, 46 (1989).

\bibitem[\protect\citeauthoryear{West}{1989}]{West1989}
M.J.~West, \ApJ, {\bf 347}, 610 (1989).

\bibitem[\protect\citeauthoryear{Catelan et al.}{2001}]{Catelanetal2001}
P.~Catelan, M.~Kamionkowski, R.D.~Blandford, \MNRAS, {\bf 320}, L7 (2001).

\bibitem[\protect\citeauthoryear{Heavens, Refregier \& Heymans}{2000}]
{Heavensetal2000} A.~Heavens, A.~Refregier, C.~Heymans, \MNRAS, {\bf 319},
649 (2000).

\bibitem[\protect\citeauthoryear{Croft \&
Metzler}{2000}]{CroftMetzler2000} R.A.C.~Croft, C.A.~Metzler, \ApJ,
{\bf 545}, 561 (2000).

\bibitem[\protect\citeauthoryear{Hopkins, Bahcall \& Bode}{2005}]
{Hopkinsetal2005} P.F.~Hopkins, N.A.~Bahcall, P.~Bode, \ApJ, {\bf
618}, 1 (2005).

\bibitem[\protect\citeauthoryear{Hahn et al.}{2007}]{Hahnetal2007}
O.~Hahn, C.~Porciani, C.C.~Carollo, A.~Dekel, \MNRAS, {\bf 375}, 489
(2007).

\bibitem[\protect\citeauthoryear{Platen et al.}{2007}]{Platenetal2008}
E.~Platen, R.~van de Weygaert, B.J.T.~Jones, ArXiv Astrophysics 
eprint:astro-ph/0711.2480 (2007).

\bibitem[\protect\citeauthoryear{Adler}{1981}]{Adler1981} R.J.~Adler, 
The Geometry of Random Fields (Chichester:Wiley, 1981).

\bibitem[\protect\citeauthoryear{Peacock \& Heavens}{1985}]
{PeacockHeavens1985} J.A.~Peacock, A.F.~Heavens, \MNRAS, {\bf 217}, 805
(1985).

\bibitem[\protect\citeauthoryear{Bardeen et al.}{1986}]{Bardeenetal1986}  
J.M.~Bardeen, J.R.~Bond, N.~Kaiser, A.S.~Szalay, \ApJ, {\bf 304}, 15  
(1986).

\bibitem[\protect\citeauthoryear{Doroshkevich}{1970}]{Doroshkevich1970}
A.G.~Doroshkevich, Astrofizika, {\bf 6}, 581 (1970).

\bibitem[\protect\citeauthoryear{Doroshkevich \& Shandarin}{1978}]
{DoroshkevichShandarin1978} A.G.~Doroshkevich, S.F.~Shandarin, 
Soviet.~Astr., {\bf 22}, 653 (1978).

\bibitem[\protect\citeauthoryear{Lee \&
Shandarin}{1998}]{LeeShandarin1998}  J.~Lee, S.~Shandarin, \ApJ, {\bf
500}, 14 (1998).

\bibitem[\protect\citeauthoryear{Catelan \& Porciani}{2001}]
{CatelanPorciani2001} P.~Catelan , C.~Porciani, \MNRAS, {\bf 323}, 713
(2001).

\bibitem[\protect\citeauthoryear{Desjacques}{2007}]{Desjacques2007}
V.~Desjacques, astro-ph/0707.4670 (2007).

\bibitem[\protect\citeauthoryear{Bernardeau}{1994}]{Bernardeau1994} 
F.~Bernardeau, \ApJ, {\bf 427}, 51 (1994).

\bibitem[\protect\citeauthoryear{Pogosyan et al.}{1998}]{Pogosyanetal1998}
D.~Pogosyan, J.R.~Bond, L.~Kofman, J.~Wadsley, 1998, in
Proceedings of the XIV IAP Colloquium, Wide Field Surveys in
Cosmology, ed. Y. Mellier and S. Colombi (Paris: Editions Frontieres).

\bibitem[\protect\citeauthoryear{}{}]{gaussianity}{
Komatsu, et al., \ApJS, {\bf 148}, 119 (2003);
P.~Creminelli, N.~Alberto, L.~Senatore, M.~Tegmark, M.~Zaldarriaga,
\JCAP, {\bf 05}, 004 (2006);
A.~Yadav, B.D.~Wandelt, ArXiv Astrophysics eprint:astro-ph/0712.1148 (2007);
E.~Komatsu, et al., ArXiv Astrophysics eprint:astro-ph/0803.0547 (2008).
\label{gaussianity}}

\bibitem[\protect\citeauthoryear{Lee \& Pen}{2001}]{LeePen2001}
J.~Lee,  U.-L.~Pen, \ApJ, {\bf 555}, 106 (2001).

\bibitem[\protect\citeauthoryear{Crittenden et
al.}{2001}]{Crittendenetal2001}  R.G.~Crittenden, N.~Priyamvada,
U.-L.~Pen, T.~Theuns, \ApJ, {\bf 559}, 552 (2001).

\bibitem[\protect\citeauthoryear{}{}]{isotropic} {$P$ contains
infinitely many $j=0$ (irreducible) representations of SO(3).
\label{isotropic}}

\bibitem[\protect\citeauthoryear{Harish Chandra}{1958}]{Harish1958}
Harish-Chandra, \AmJP, {\bf 80}, 241 (1958).

\bibitem[\protect\citeauthoryear{Itzykson \&
Zuber}{1980}]{ItzyksonZuber1980}  C.~Itzykson, J.B.~Zuber J.B., \JMP,
{\bf 21}, 411 (1980).

\bibitem[\protect\citeauthoryear{Muirhead}{1982}]{Muirhead1982}
R.J.~Muirhead, {\it Aspects of Multivariate Statistical Theory}
(Wiley, New-York, 1986)\;.

\bibitem[\protect\citeauthoryear{}{}]{fastcomp}
{P.~Koev and A.~Edelman, Math.~Comp.~, {\bf 75}, 833 (2006) for an 
algorithm linear in the size of the matrix arguments.
\label{fastcomp}}

\bibitem[\protect\citeauthoryear{Berntsen, Espelid \& Genz}{1991}]
{Berntsenetal1991} J.~Berntsen, T.O.~Espelid, A.~Genz, ACM
Transactions on Mathematical Software, {\bf 17}, 437 (1991).

\bibitem[\protect\citeauthoryear{}{}]{haloloc}{
S.F.~Shandarin, A.A.~Klypin, Soviet~Astron., {\bf 28}, 491 (1984);
C.S.~Frenk, S.D.M.~White, M.~Davis, G.~Efstathiou, \ApJ, {\bf 327},
507 (1988);
N.~Katz, T.~Quinn, J.M.~Gelb, \MNRAS, {\bf 265}, 689 (1993);
C.~Porciani, A.~Dekel, Y.~Hoffman, \MNRAS, {\bf 332},  339 (2002).
\label{haloloc}}

\bibitem[\protect\citeauthoryear{Kaiser}{1984}]{Kaiser1984} N.~Kaiser,
\ApJ, {\bf 284}, L9 (1984).

\bibitem[\protect\citeauthoryear{Pen et al.}{2001}]{Penetal2001}
U.L.~Pen, J.~Lee, U.~Seljak, \ApJ, {\bf 543}, L107 (2000).

\bibitem[\protect\citeauthoryear{Lee \& Pen}{2007}]{LeePen2007}
J.~Lee, U.-L.~Pen, ArXiv Astrophysics eprint:astro-ph/0707.1690 
(2007).

\bibitem[\protect\citeauthoryear{Hirata \&
Seljak}{2004}]{HirataSeljak2004} C.M.~Hirata, U.~Seljak, \PRD, {\bf
70}, 063526 (2004).

\bibitem[\protect\citeauthoryear{}{}]{align}{
R.E.~Smith, P.I.R.~Watts, \MNRAS, {\bf 360}, 203 (2005);
R.E.~Smith, P.I.R.~Watts, R.K.~Sheth, \MNRAS, {\bf 365}, 214 (2006).
\label{align}}

\bibitem[\protect\citeauthoryear{Matarrese, verde \& Jimenez}{2000}]
{Matarreseetal2000} S.~Matarrese, L.~Verde, R.~Jimenez, \ApJ, {\bf 541},
10 (2000).

\bibitem[\protect\citeauthoryear{Komatsu \& Spergel}{2001}]
{KomatsuSpergel2001} E.~Komatsu, D.N.~Spergel, \PRD, {\bf 63}, 063002 
(2001). 

\bibitem[\protect\citeauthoryear{Maldacena}{2003}]{Maldacena2003}
J.~Maldacena, \JHEP, {\bf 5}, 13 (2003).

\bibitem[\protect\citeauthoryear{Hui \& Zhang}{2002}]{HuiZhang2002}
L.~Hui, Z.~Zhang, ArXiv Astrophysics eprint:astro-ph/0205512.

\bibitem[\protect\citeauthoryear{Bond \& Myers}{1996}]{BondMyers1996}
J.R.~Bond, S.T.~Myers, \ApJS, {\bf 103}, 1 (1996).

\bibitem[\protect\citeauthoryear{}{}]{zetalimit}  {$\zeta_{11}$ is
always less than $(13\pi-27)/30\pi\approx 0.147$, a value reached
only when $R_1=R_2$.}

\end{thebibliography}
\end{document}